\documentclass[a4paper]{revtex4}
\usepackage{graphicx}
\usepackage{amsmath}
\usepackage{amssymb}
\usepackage{amsfonts}

\newcommand{\be}{\begin{equation}}
\newcommand{\ee}{\end{equation}}
\newcommand{\bea}{\begin{eqnarray}}
\newcommand{\eea}{\end{eqnarray}}
\newcommand{\beaa}{\begin{eqnarray*}}
\newcommand{\eeaa}{\end{eqnarray*}}

\newcommand{\e}{\mathrm{e}}

\begin{document}

\title{Cosmological solutions in F(R) Ho\v{r}ava-Lifshitz gravity} 

\author{Diego S\'{a}ez-G\'{o}mez}

\address{Consejo Superior de Investigaciones Cient\'{\i}ficas,
ICE(CSIC-IEEC), Campus UAB Facultat de Ci\`encies, Torre C5-Parell-2a
pl, E-08193 Bellaterra (Barcelona) Spain}

\begin{abstract}
 At the present work, it is studied the extension of $F(R)$ gravities to the new recently proposed theory of gravity, the so-called Ho\v{r}ava-Lifshitz gravity, which provides a way to make the theory power counting renormalizable by breaking Lorentz invariance. It is showed that dark energy can be well explained in the frame of this extension, just in terms of gravity. It is also explored the possibility to unify inflation and late-time acceleration under the same mechanism, providing a natural explanation the accelerated expansion. 
\end{abstract}


\maketitle

\section{Introduction to F(R) Ho\v{r}ava-Lifshitz gravity}

Since observational data suggests the well known fact that probably the Universe is immersed in an accelerating phase, a lot of models have been proposed in order to explain such unusual behavior of gravity at large scales. The first and still most popular model was the cosmological constant, which gives a negative pressure and then, an acceleration on the expansion. Nevertheless, the fine tuning problem on the model has made that another kind of candidates has been seriously taken into account. One of them suggests the possibility of the modification of  gravitational field equations such that at large scales, gravity behaves different than in local systems, producing  accelerated expansion and even open the possibility to explain, at the same time, the inflationary epoch in a natural way. One of this theories is the so-called $F(R)$ gravity, which just generalized the Hilbert-Einstein action to more complex ones (for a recent review see Ref.~\cite{FRreview}).
On the other hand, recently a new gravitational theory has emerged, which suggests to be power-counting renormalizable but breaking explicitly Lorentz invariance \cite{Horava} by means of different scaling properties on the coordinates,
\be
x^i=b x^i\, , \quad t=b^zt\, .
\label{1.1}
\ee
where $z$ is a dynamical critical exponent that makes the theory to be renormalizable for $z=3$ in a $3+1$ spacetime dimensions. This new theory, already known as Ho\v{r}ava-Lifshitz gravity,  does not explain the dark energy epoch in its original form as it also occurs in General Relativity (GR). Then, generalizations of the original action has been suggested in the same way as $F(R)$ gravity from GR in order to explain the dark energy and even the inflationary epochs under just a gravitational mechanism. Let us briefly review this extension called $F(R)$ Ho\v{r}ava-Lifshitz gravity. Giving a general metric written in the ADM decomposition in a 3+1 spacetime dimensions,
\be
ds^2=-N^2 dt^2+g^{(3)}_{ij}(dx^i+N^idt)(dx^j+N^jdt)\, ,
\label{1.2}
\ee 
The action considered here can be written as \cite{FRhorava},
\[
S=\frac{1}{2\kappa^2}\int dtd^3x\sqrt{g^{(3)}}N F(\tilde{R})\, , 
\]
\be
\tilde{R}= K_{ij}K^{ij}-\lambda K^2 + R^{(3)}+
2\mu\nabla_{\mu}(n^{\mu}\nabla_{\nu}n^{\nu}-n^{\nu}\nabla_{\nu}n^{\mu})-
L^{(3)}(g_{ij}^{(3)})\, ,
\label{1.3}
\ee
where $K_{ij}$ is the extrinsic curvature, and $R^{(3)}$ is the spatial scalar curvature. Note that, apart of the gravitational coupling $\kappa$, there are  two new constants $(\lambda,mu)$, which account for the violation of the full diffeomorphism. Original Ho\v{r}ava-Lifshitz gravity is given by $F(\tilde{R})=\tilde{R}$, where the term in front of $\mu$ becomes a total derivative and is omitted. The term $L^{(3)}(g_{ij}^{(3)})$ is chosen to be $L^{(3)}(g_{ij}^{(3)})=E^{ij}G_{ijkl}E^{kl}$ with $G_{ijkl}$ is the generalized De Witt metric, and the expression for $E_{ij}$ is constructed to
satisfy the ``detailed balance principle'', being  defined through variation of an action (for more details see Ref.~\cite{Horava}). \\
Here, we are interested to study solutions of a spatially flat FRW Universe, described by the metric,
\be
ds^2=-N^2dt^2+a^2(t)\sum_{i=1}^3 \left(dx^{i}\right)^2\, .
\label{1.4}
\ee
If we also assume the projectability condition,
$N$ can be taken to be just time-dependent and, by using the
foliation-preserving
diffeomorphisms, it can be fixed to be unity, $N=1$. 
For the metric (\ref{1.4}), the scalar $\tilde{R}$ is
given by $\tilde{R}=\frac{3(1-3\lambda
+6\mu)H^2}{N^2}+\frac{6\mu}{N}\frac{d}{dt}\left(\frac{H}{N}\right)$, and assuming the FRW metric (\ref{1.4}),
the second FRW equation can be obtained by varying
the action with respect to the spatial metric $g_{ij}^{(3)}$, which
yields
\[
0=F(\tilde{R})-2(1-3\lambda+3\mu)\left(\dot{H}+3H^2\right)F'(\tilde{R})-2(1-
3\lambda)
\dot{\tilde{R}}F''(\tilde{R})
\]
\be
+2\mu\left(\dot{\tilde{R}}^2F^{(3)}(\tilde{R})+\ddot{\tilde{R}}F''(\tilde{R})\right)+\kappa^2p_m\, ,
\label{1.6}
\ee
here $\kappa^2=16\pi G$, $p_m$ is the pressure of a perfect fluid
that fills the Universe, and $N=1$. Note that this
equation becomes the usual second FRW equation for convenient $F(R)$
gravity (\ref{1.4}), by setting the constants $\lambda=\mu=1$.
Assuming the projectability condition,
variation over $N$ of the action (\ref{1.3}) yields the first FRW
 equation
\be
0=F(\tilde{R})-6\left[(1-3\lambda
+3\mu)H^2+\mu\dot{H}\right]F'(\tilde{R})+6\mu H
\dot{\tilde{R}}F''(\tilde{R})-\kappa^2\rho_m\, ,
\label{1.8}
\ee
Hence, starting from a given $F(\tilde{R})$ function, and solving 
Eqs.~(\ref{1.6}) and (\ref{1.8}), a cosmological solution
can be obtained.

\section{$\Lambda$CDM model in F(R) Ho\v{r}ava-Lifshitz gravity}

Let us now discuss some cosmological solutions of
$F(\tilde{R})$ Ho\v{r}ava-Lifshitz gravity, specially the solution that reproduces the $\Lambda$CDM model. The first FRW equation,
given by (\ref{1.8}) with $C=0$, can be rewritten as a function of
the number of e-foldings $\eta=\ln\frac{a}{a_0}$, instead of the usual
time $t$. This technique has been developed in Ref.~\cite{FR1} for
convenient $F(R)$ gravity, where it was shown that any $F(R)$
theory can be reconstructed for a given cosmological solution. Here,
we extend such formalism to the Ho\v{r}ava-Lifshitz $F(R)$ gravity. By using the energy conservation
equation, and assuming a perfect fluid with equation of state (EoS)
$p_m=w_m\rho_m$, the energy density yields $\rho_m=\rho_0 a^{-3(1+w_m)}=\rho_0a_0^{-3(1+w_m)}\e^{-3(1+w_m)\eta}$. And since $\frac{d}{dt}=H\frac{d}{d\eta}$ and $\frac{d^2}{d\eta^2}
=H^2\frac{d^2}{d\eta^2}+H\frac{dH}{d\eta}\frac{d}{d\eta}$,
the FRW eq.~(\ref{1.8}) takes the form
\[
0=F(\tilde{R})-6\left[(1-3\lambda+3\mu)G+\frac{\mu}{2}G'\right]\frac{dF(\tilde{R})
}{d\tilde{R}}+18\mu\left[(1-3\lambda+6\mu)GG'+\mu GG''\right]
\frac{d^2F(\tilde{R})}{d^2\tilde{R}}
\]
\be
-\kappa^2\rho_0a_0^{-3(1+w)}
\e^{-3(1+w)\eta}\, ,
\label{2.1}
\ee
where the primes denote derivatives with
respect to $\eta$, and we have rewritten the Hubble parameter as $G(\eta)=H^2(\eta)$. The scalar curvature is now written as
$\tilde{R}=3(1-3\lambda+6\mu)G+3\mu
G'$.  Hence,
for a given cosmological solution $H^2=G(\eta)$, one can resolve
Eq.~(\ref{2.1}), and the $F(\tilde{R})$ that reproduces such solution
is obtained. \\
Let us consider the solution that reproduces the $\Lambda$CDM model. In this case the Hubble parameter takes the form,
\be
H^2 =G(\eta)= H_0^2 + \frac{\kappa^2}{3}\rho_0 a^{-3} = H_0^2 +
\frac{\kappa^2}{3}\rho_0 a_0^{-3} \e^{-3\eta} \, .
\label{2.2}
\ee
Then, by introducing the solution (\ref{2.2}) in the equation (\ref{2.1}), and by performing a change of variable $x=\frac{\tilde{R}-9\mu H^2_0}{3H^2_0(1+3(\mu-\lambda))}$, and after some algebraic manipulation, it yields the differential equation,
\be
0=x(1-x)\frac{d^2 F}{dx^2} + \left(\gamma - \left(\alpha + \beta +
1\right)x\right)\frac{dF}{dx} - \alpha \beta F-3H^2_0(1+3(\mu-\lambda))x-3(1-3\lambda+9\mu)H^2_0\, ,
\label{2.3}
\ee 
with the set of parameters $(\alpha,\beta,\gamma)$ being given by
\be
\gamma=-\frac{1}{2}\, , \quad \alpha+\beta=
\frac{1-3\lambda-\frac{3}{2}\mu}{3\mu}\, , \quad
\alpha\beta=-\frac{1+3(\mu-\lambda)}{6\mu}\, .
\label{2.4}
\ee
The complete solution of Eq.~(\ref{2.3}) is a Gauss' hypergeometric function
plus a linear term and a cosmological constant coming, namely
\be
F(\tilde{R}) = C_1 F(\alpha,\beta,\gamma;x) + C_2 x^{1-\gamma} F(\alpha -
\gamma + 1, \beta - \gamma +
1,2-\gamma;x)+\frac{1}{\kappa_1}\tilde{R}-2\Lambda\, ,
\label{D11}
\ee
where $C_1$ and $C_2$ are constants, $\kappa_1=3\lambda-1$ and
$\Lambda=-\frac{3H_0^2(1-3\lambda+9\mu)}{2(1-3\lambda+3\mu)}$. In this case, the solution (\ref{D11}) behaves similarly
to the classical $F(R)$ theory, except that now the parameters of
the theory depend on $(\lambda,\mu)$, which are
allowed to vary.
We can conclude that the cosmic evolution
described by the $\Lambda$CDM model given by the Hubble parameter (\ref{2.2}) is well reproduced by this class of
theories. Note that for an special choice of the parameters $\mu=\lambda-\frac{1}{3}$, which decouples the scalar mode included in all $F(\tilde{R})$ models from the metric tensor (for more details see Ref.~\cite{FRhorava}), the solution reduces to $F(\tilde{R})=\frac{1}{\kappa_1}\tilde{R}-2\Lambda$, which is equivalent to a cosmological constant. \\
Nevertheless, other class of solutions for dark energy (as phantom models) can well be reproduced in this class of theories as it was showed in Ref.~\cite{FRHorava2}.

\section{Unifying inflation with dark energy}

Here we
extend the  class of
models called viable $F(R)$ gravity, which pass the local gravity tests, and can reproduce dark energy and even inflation, to the Ho\v{r}ava-Lifshitz gravity. We consider the action, $F(\tilde{R})=\tilde{R}+f(\tilde{R})$, assuming that the term $f(\tilde{R})$ becomes important at
cosmological scales, while for scales compared with the Solar system one
the theory becomes linear on $\tilde{R}$. As an example, we consider
the following function, 
\be
f(\tilde{R})=\frac{\tilde{R}^n(\alpha\tilde{R}^n-\beta)}{1+\gamma
\tilde{R}^n}\, ,
\label{3.1}
\ee
where $(\alpha, \beta, \gamma)$ are constants and $n>1$. This theory
reproduces the inflationary and cosmic acceleration epochs in standard
$F(R)$ gravity, which  is also the case  in the
present theory, as will be shown. During inflation, it is assumed
that the curvature scalar tends to infinity. In this case the model
(\ref{3.1}),  behaves as $\lim_{\tilde{R}\rightarrow\infty}F(\tilde{R})=\alpha\tilde{R}^n$. 
Then, by solving the FRW equation (\ref{1.8}), this kind of function
yields a power-law solution of the type
\be
H(t)=\frac{h_1}{t}\, , \quad \text{where} \quad
h_1=\frac{2\mu(n-1)(2n-1)}{1-3\lambda+6\mu-2n(1-3\lambda+3\mu)}\, .
\label{3.4}
\ee
This solution produces acceleration during the inflationary epoch if
the parameters of the theory are properly defined. The acceleration
parameter is given by $\frac{\ddot{a}}{a}=h_1(h_1-1)/t^2$, thus, for
$h_1>1$ the inflationary epoch is well reproduced by the model
(\ref{3.1}). On the other hand, the function (\ref{3.1}) has a
minimum at $\tilde{R}_0$, given by
\be
\tilde{R}_0 \sim \left( \frac{\beta}{\alpha\gamma}\right)^{1/4}\, ,
\qquad f'(\tilde{R})=0\, , \qquad  f(\tilde{R})= -2\Lambda\sim
-\frac{\beta}{\gamma}\, ,
\label{3.5}
\ee
where we have imposed the condition $\beta\gamma/\alpha\gg 1$. Then, at
the current epoch the scalar curvature acquires a small value which
can be fixed to coincide with the minimum (\ref{3.5}), such that the
FRW equations (\ref{1.6}) and (\ref{1.8}) yield
\be
H^2=\frac{\kappa^2}{3(3\lambda-1)}\rho_m+\frac{2\Lambda}{3(3\lambda-1)}\, 
\quad \dot{H}=-\kappa^2\frac{\rho_m+p_m}{3\lambda-1}\, ,
\label{3.6}
\ee
which look very similar to the standard FRW equations in GR, except
for the parameter $\lambda$. As at the current
epoch the scalar $\tilde{R}$ is  small, the theory is
in the IR limit where the parameter $\lambda\sim1$, and the equations
approach the usual ones for $F(R)$ gravity. Hence, the FRW
equations (\ref{3.6}) reproduce the behavior of the well known
$\Lambda$CDM model with no need to introduce a dark energy fluid
to explain the current universe acceleration, as well as the inflationary epoch.  \\

Hence, we can conclude that the extension  $F(R)$  Ho\v{r}ava-Lifshitz gravity can well reproduce the dark energy and inflationary epochs with no need of any exotic field. It has been showed that viable models as (\ref{3.1}) can be extended to this new class of gravity, and adjusted to reproduce the whole expansion history. However, a deeper analysis has to be performed in order to obtain more information on this class of theories (see Ref.~\cite{DSG}). Even the problems on the scalar graviton can be resolved by introducing a U(1) symmetry in  $F(R)$ Ho\v{r}ava-Lifshitz gravity (see Ref.~\cite{FR_HL_U1}). 

\section*{Acknowledgements} I would like to thank Emilio Elizalde, Sergei Odintsov and Josef Kluso\v{n} for useful discussions. I acknowledge an FPI grant, project FIS2006-02842, provided by Ministry of Science (Spain).

\end{document}